\documentclass[11pt,a4paper]{article}
\usepackage[left=15mm, top=10mm, right=15mm, bottom=20mm, nohead=true, nofoot=true]{geometry}

\usepackage[utf8]{inputenc}
\usepackage{authblk}
\usepackage{indentfirst}
\usepackage{misccorr}
\usepackage{graphicx}
\usepackage{amsmath}
\usepackage{amssymb}
\usepackage{multicol}
\usepackage{bm}
\usepackage{longtable}
\usepackage{pdflscape}
\usepackage{epstopdf}
\usepackage{multirow}
\usepackage{adjustbox}
\usepackage{amsfonts}
\usepackage{ifthen}
\usepackage{fancyhdr}
\usepackage{setspace}
\usepackage{xcolor}
\usepackage[round,authoryear,comma,sort&compress,numbers]{natbib}
\usepackage[toc,title,page]{appendix}
\usepackage[hidelinks]{hyperref}
\usepackage{url}

\hypersetup{
    colorlinks=true,
    linkcolor=green,
    citecolor=blue,
    filecolor=magenta,
    urlcolor=cyan,
    pdftitle={Sharelatex Example},
    pdfpagemode=FullScreen,
}

\fancypagestyle{plain}{\chead{\texttt{\textcolor{brown}{Communications of BAO, Vol. 72, Issue 2, 2025, pp. ???-???}}}}
\title{\textbf{Half-Light Radius Measurements of Andromeda Dwarf Satellites from the Isaac Newton Telescope Survey Using Exponential, Plummer, and Sérsic Fits}}
\author[1, 2]{Hedieh Abdollahi \thanks{hedieh.abdollahi@csfk.org}}

\affil[1]{\scriptsize Konkoly Observatory, HUN-REN Research Centre for Astronomy and Earth Sciences, MTA Centre of Excellence, Konkoly-Thege Mikl\'os \'ut 15-17, H-1121, Budapest, Hungary}
\affil[2]{\scriptsize School of Astronomy, Institute for Research in Fundamental Sciences (IPM), P.O. Box 1956836613, Tehran, Iran}

\def\apj{Astrophys.~J. }

\def\pasp{Publ.~Astron.~Soc.~Pac. }

\def\aap{Astron.~Astrophys. }

\def\apjl{Astrophys.~J.~Lett. }

\def\aj{Astron.~J. }
\def\mnras{Mon.~Not.~R.~Astron.~Soc. }

\def\aapr{Astron.~Astrophys.~Rev. }

\begin{document}
\pagestyle{empty}
\newpage
\pagestyle{fancy}
\label{firstpage}
\date{}
\maketitle

\begin{abstract}
We present half-light radius measurements for the dwarf satellites of Andromeda, based on multi-epoch imaging from the Isaac Newton Telescope (INT) Monitoring Survey of Local Group dwarf galaxies. This analysis is conducted within a larger study to identify long-period variable (LPV) stars in these galaxies. The survey was performed with the Wide Field Camera on the 2.5-m INT and covers multiple epochs obtained between 2015 and 2018 in the $i$ (Sloan) and $V$ (Harris) bands. To determine the half-light radii, we derived surface brightness and number density profiles for each system and fitted them with Exponential, Plummer, and S$\acute{e}$rsic models. The resulting half-light radii are in good agreement with literature values but reveal subtle variations linked to differences in stellar distribution and morphology. Distances were independently estimated using the Tip of the Red Giant Branch (TRGB) method, yielding values consistent with previous determinations. The complete photometric and variability catalogs will be made publicly available through CDS/VizieR, providing a valuable resource and foundation for future studies of the structure, stellar populations, and evolution of Andromeda’s dwarf companions.

\end{abstract}

\emph{\textbf{Keywords:} stars: AGB and LPV – stars: evolution – stars: mass-loss – galaxies: dwarf – galaxies: Local Group – surveys}

\section{Introduction}
Dwarf galaxies in the Local Group serve as crucial laboratories for studying stellar evolution and galaxy formation across a broad range of masses, metallicities, and environments \citep{2011MNRAS.411..263J, 2023ApJ...948...63A}. Their proximity allows their stellar populations to be individually resolved, enabling detailed reconstructions of star-formation histories (SFHs), chemical enrichment processes, and the effects of feedback in shaping low-mass galaxies \citep{2011MNRAS.411..263J, 2021ApJ...923..164S}. As the most numerous type of galaxy in the Universe, dwarfs are thought to be the building blocks of larger systems, offering valuable insights into hierarchical galaxy assembly and the baryon cycle \citep{2023ApJ...948...63A, Mahani25}.

Among their evolved stellar populations, asymptotic giant branch (AGB) stars play a key role in tracing the final stages of stellar evolution and the recycling of material into the interstellar medium through mass-loss and dust production \citep{2017ApJ...851..152B, Hoefner18, 2025ApJ...991...24B}. Within this population, long-period variables (pulsating AGB stars with large amplitudes and periods ranging from a few hundred to over a thousand days) are particularly powerful probes of late stellar evolution \citep{2018AJ....156..112Y, 2019ApJ...877...49G, 2023ApJ...942...33P}. Their variability arises from thermal pulsations during the advanced stages of evolution, and their brightness makes them detectable even in faint, low-metallicity dwarf systems across the Local Group \citep{Whitelock23}.

LPVs provide valuable diagnostics of stellar populations, as their luminosities are closely linked to their initial masses and, therefore, to the ages of their parent populations \citep{vanLoon08}. Their spatial and luminosity distributions thus encode information about a galaxy’s star-formation and chemical enrichment histories, while their variability properties offer direct constraints on mass-loss and dust feedback in metal-poor environments. Because of these characteristics, LPVs are ideal tracers of intermediate-age and old stellar populations in dwarf galaxies, bridging the gap between the short-lived massive stars and the long-lived red giant branch (RGB) stars \citep{2021A&A...647A.170L}.

In this study, we investigate the populations of LPVs in 17 satellite galaxies of Andromeda (M31). By identifying and characterizing these evolved stars, we aim to reconstruct the star-formation histories of the host galaxies and to quantify the dust production and mass-loss rates associated with LPVs in low-metallicity environments. Through the analysis of their spatial distributions, luminosities, and pulsation properties, we also constrain the structural parameters and evolutionary pathways of M31’s satellite system. This work complements previous studies of LPVs in Milky Way companions and isolated dwarf galaxies \citep[e.g.,][]{2011MNRAS.411..263J, 2021ApJ...923..164S, 2021ApJ...910..127N, 2023ApJ...948...63A, 2023ApJ...942...33P,  mahani23, Gholami25}. The observations and data reduction procedures are described in Section~\ref{sec:observations}. The method used for identifying LPVs is outlined in Section~\ref{sec:Identification}, the results are presented in Section~\ref{sec:Results}, and the conclusions are summarized in Section~\ref{sec:Conclusions}.

\section{Observations and Data}
\label{sec:observations}
Observations were obtained as part of the Isaac Newton Telescope Monitoring Survey of Local Group Dwarf Galaxies, designed to explore stellar variability in 55 dwarf galaxies and four globular clusters. The survey provides multi-epoch optical photometry in the $V$ (Harris) and $i$ (Sloan) bands, with sensitivity sufficient to detect AGB stars near the tip of the red giant branch \citep{2021ApJ...923..164S}. 

All data were collected using the Wide Field Camera (WFC) mounted on the 2.5-m INT at the Observatorio del Roque de los Muchachos, La Palma. The WFC consists of four 2048$\times$4096 CCDs with a pixel scale of 0.33\,arcsec\,pixel$^{-1}$, providing a total field of view of 34$\times$34 arcmin$^2$. Each target was observed in nine epochs between June 2015 and February 2018 through the $i$ and $V$ filters, with an additional $I$-band frame obtained during several of the early epochs.

Data reduction followed standard procedures using the {\sc \texttt{THELI}} pipeline, including bias subtraction, dark and flat-field correction, astrometric alignment, and stacking \citep{Erben05}. Point-spread function (PSF) photometry was performed using the {\sc \texttt{DAOPHOT/ALLFRAME}} suite \citep{Stetson94}, optimized for crowded stellar fields. Instrumental magnitudes were transformed to the standard system using Landolt standard stars \citep{Landolt92}, and aperture corrections were derived from 30–40 bright, isolated stars in each frame.

Photometric completeness was assessed through artificial-star experiments using the {\sc \texttt{ADDSTAR}} routine. The resulting completeness is $\geq90$\% down to $i,V\sim22$ mag and approximately 50\% at 23 mag, ensuring coverage of the AGB population up to the TRGB. The observational parameters, including the observing cadence, filters, exposure times, and image quality, follow those described in detail by \citet{2020ApJ...894..135S}.

\section{Identification of Long-Period Variables}
\label{sec:Identification}
The identification of  LPVs was performed using the variability indices introduced by \citet{Welch93} and refined by \citet{Stetson96}. These indices quantify the degree of correlated brightness variation across multiple epochs and filters, providing a robust statistical method for detecting variable stars even when light-curves are sparsely sampled. The approach combines three key components:
\begin{itemize}
\item {\bf J index:} measures the correlation between magnitude deviations in the $i$ and $V$ bands for each star across epochs;
\item {\bf K index:} compares the mean absolute deviation of the magnitudes to their root-mean-square deviation, providing a measure of the light-curve kurtosis and thus the shape or sharpness of the observed variability;
\item {\bf L index:} represents a weighted combination of $J$ and $K$, optimized to enhance the contrast between variable and non-variable stars.
\end{itemize}

Stars with $L$ indices exceeding the expected Gaussian noise distribution were flagged as candidate variables. Following \citet{Stetson96}, we adopted a conservative detection threshold of $L \geq 1.0$ to ensure reliable identification across fields of varying depth and crowding. For each candidate, the variability amplitude was estimated assuming an approximately sinusoidal light-curve, following
\begin{equation}
A = \frac{2\sigma}{0.707},
\label{eq:Amp_i}
\end{equation}
where $\sigma$ is the standard deviation of the measured magnitudes. Only sources exhibiting amplitudes larger than 0.2 mag in the $i$ band were retained as confirmed LPV candidates, corresponding to the typical variability range expected for pulsating AGB stars near or above the tip of the red giant branch \citep{Abdollahi24}.

Foreground contamination was mitigated through cross-matching with {\it Gaia}~DR3 \citep{GaiaDR3}. Sources showing significant parallaxes or proper motions inconsistent with M31’s systemic motion were excluded, using the criteria:
\[
\sqrt{(\mu_{\alpha})^2 + (\mu_{\delta})^2} > 0.28~\mathrm{mas\,yr^{-1}} + 2.0\,\mathrm{error},
\]
or $\pi / \sigma_\pi \geq 2$, where $\pi$ is the measured parallax and $\sigma_\pi$ its uncertainty. This procedure effectively removes Galactic foreground stars and ensures that the remaining variables are consistent with the distances of M31’s satellite galaxies \citep{2023ApJ...942...33P}.

This statistical method provides a consistent and objective means of identifying LPVs in sparsely sampled, multi-epoch data sets. Unlike classical period-finding techniques, which require densely sampled light-curves, the Welch–Stetson index approach is optimized for surveys with limited temporal coverage, such as the INT Monitoring Survey. The resulting LPV catalog forms the basis for subsequent analyses of the spatial distribution, variability properties, and dust-producing AGB populations within the M31 satellite system \citep{2023ApJ...948...63A, Mahani25, SEDust25}.

\section{Results}
\label{sec:Results}

We estimated several physical parameters of the LPV populations associated with each galaxy. Using multi-band photometric data, we derived the TRGB, distance modulus, and half-light radius by fitting three different profile functions. These measurements offer valuable insights into the late stages of stellar evolution and highlight the role of LPVs in driving dust production and chemical enrichment within their host galaxies. 

\subsection{Tip of the Red Giant Branch}

To ensure reliable distance estimates, we first constructed the $i$-band luminosity function (LF) of evolved stars for each galaxy, selecting stars within two half-light radii to minimize contamination from foreground and background sources. The LF was smoothed and then analyzed using a Sobel edge-detection filter, which highlights the sharp discontinuity in the stellar counts corresponding to the TRGB \citep{Lee93}. The TRGB position was identified as the peak in the filter response, and its apparent magnitude was measured by fitting a Gaussian to this feature. This procedure provides a robust, model-independent estimate of the TRGB magnitude, which we then used to derive the distance modulus for each galaxy. Distance moduli were obtained using the following equation:

\begin{equation}
(m-M)_0 = i_{TRGB} - M_{TRGB} - A_i
\end{equation}
 where $ A_i$ represents the Galactic extinction. Derived moduli span 23.57$\pm$0.08 to 25.62$\pm$0.17 mag, corresponding to distances of 467.74$^{+42.76}_{-39.19}$–1153.45$^{+32.31}_{-31.43}$ kpc, in good agreement with \cite{2012ApJ...758...11C}.

\subsection{Distances Moduli}
\begin{figure}[ht!]
  \centering
  \includegraphics[width=0.48\textwidth]{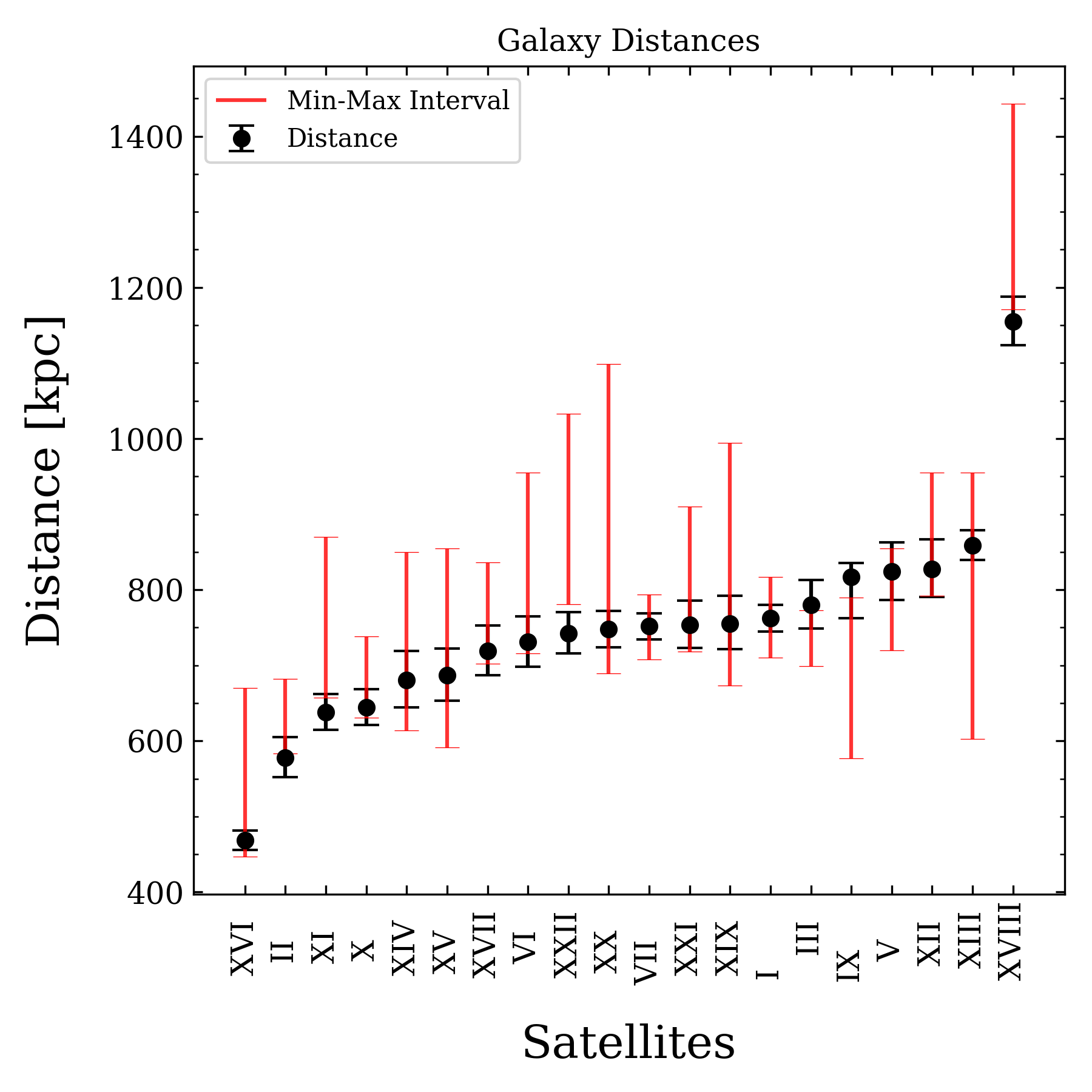}
  \caption{Distance results for Andromeda's satellite galaxies.
  The range depicted in this plot is derived from \citet{2012ApJ...758...11C}.}
  \label{fig:distance}
\end{figure}

The distances to the target galaxies were determined from the apparent magnitudes of the tip of the red giant branch in the $i$-band. As shown in Figure~\ref{fig:distance}, the derived TRGB distance moduli for the sample are in excellent agreement with independent measurements based on RR~Lyrae and Cepheid variables, generally within the expected uncertainties of $\sim$0.1--0.2~mag. This consistency confirms the reliability of our photometric calibration and validates the adopted reddening corrections.

The derived half-light radii and ellipticities further illustrate the structural diversity among the M31 satellites, ranging from compact spheroidal systems to more extended, diffuse galaxies. Such variation reflects differences in their evolutionary histories, internal dynamics, and environmental interactions within the M31 halo.

\subsection{Half-light Radius}
To measure the half-light radius of each dwarf galaxy, we analyzed how the stellar number density and surface brightness change with distance from the galaxy center. In general, both quantities decrease as one moves outward from the center. However, this relation is not always perfectly proportional, since variations in the spatial distribution of bright and faint stars can locally affect the surface brightness profile.

The stellar sample for each galaxy was first sorted by radial distance from the center. We then divided the stars into bins containing an equal number of sources and calculated both the surface brightness and the number density for each bin.

\begin{figure}[ht!]
  \centering
  \begin{minipage}{0.48\textwidth}
    \centering
    \includegraphics[width=\linewidth]{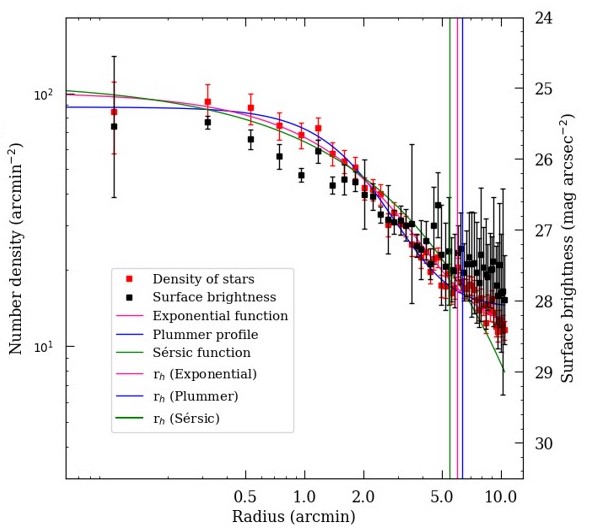}
    \caption{Comparison of half-light radius measurements of And~II, derived from three different fitting functions.}
    \label{fig:And2_rh}
  \end{minipage}
  \hfill
  \begin{minipage}{0.48\textwidth}
    \centering
    \includegraphics[width=\linewidth]{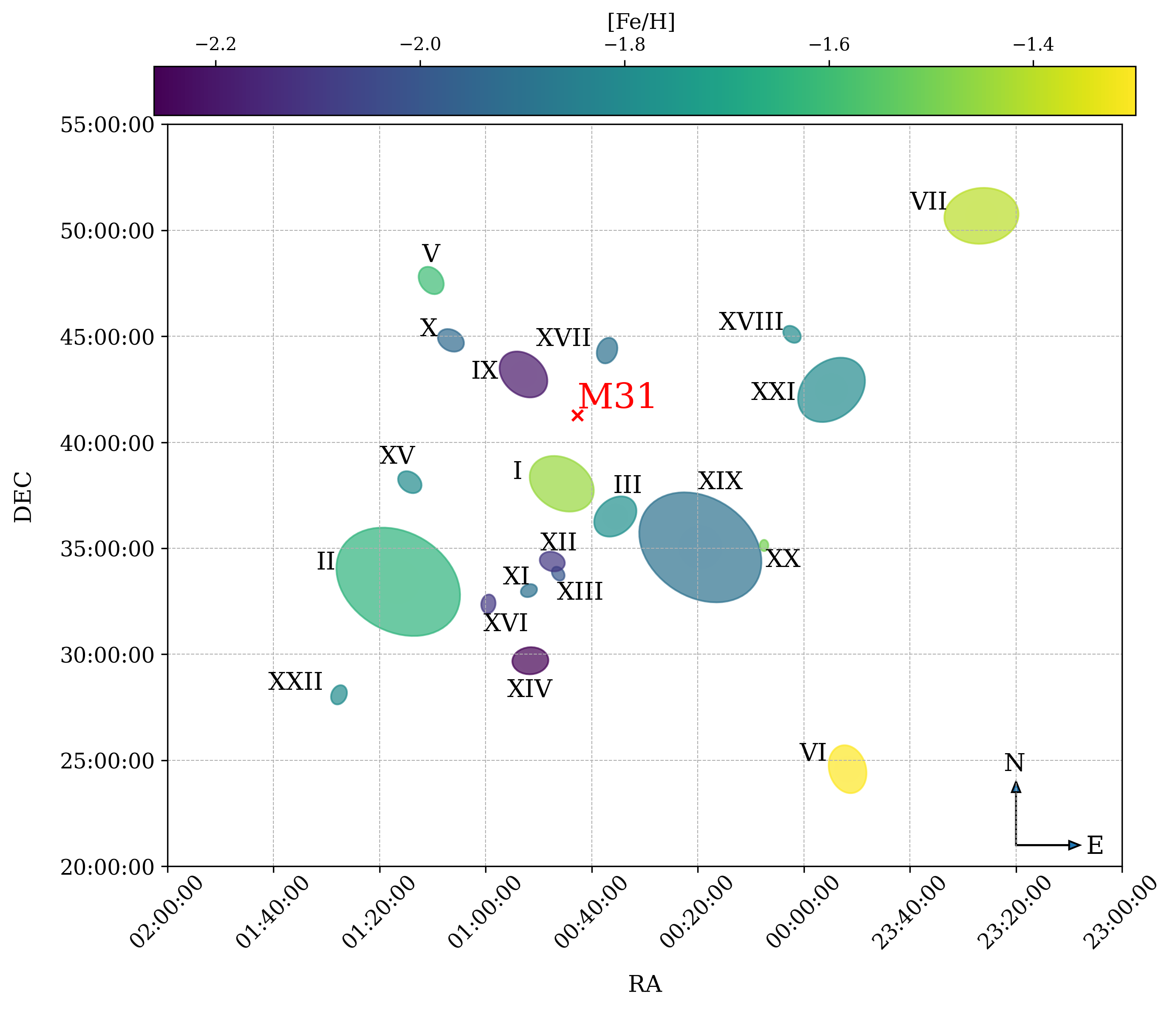}
    \caption{Spatial distribution of Andromeda’s dwarf satellite galaxies. The semi-major axis of the galaxies is represented proportionally to their actual sizes based on data from Table \ref{table:objects}.}
    \label{fig:M31_map}
  \end{minipage}
\end{figure}

Following the approach of \citet{2006MNRAS.365.1263M}, we fitted the radial profiles with several analytic models, including the Exponential (equation~\ref{equ: exp}) and Plummer (equation~\ref{equ: Plummer}) functions \citep{Faber83, Plummer11}. The King profile \citep{king66} was also tested but yielded results that deviated significantly from those obtained with the Exponential and Plummer models. To achieve more consistent and accurate estimates, we adopted the S$\acute{e}$rsic profile (equation~\ref{equ: Sersic}) \citep{Sersic68}, which provided the best overall fits to the data. Figure~\ref{fig:And2_rh} shows the half-light radius estimates for And II as an example of our sample of dwarf galaxies obtained by fitting three different profile functions. 

\begin{equation}
\label{equ: exp}
I(x) = a,e^{-x/b} + c
\end{equation}

\begin{equation}
\label{equ: Plummer}
I(x) = a,\frac{b^2}{(b^2 + x^2)^2} + c
\end{equation}

\begin{equation}
\label{equ: Sersic}
I(x) = a,e^{-b\left[\left(\frac{x}{c}\right)^{1/n} - 1\right]}
\end{equation}

The resulting half-light radii derived from the Exponential, Plummer, and S$\acute{e}$rsic fits are listed in Table~\ref{table:objects}. For comparison, we also include published values from \citet{2016ApJ...833..167M} and \citet{2012AJ....144....4M}.

Figure~\ref{fig:M31_map} presents the projected distribution of satellite galaxies around M31. The ellipses are scaled to their half-light radii derived in this study, with colors indicating metallicities from \citet{2012AJ....144....4M}. A red cross highlights the central position of M31.

\begin{table*}
\caption{Observational properties and half-light radius of targets.}
\small 
\setlength{\tabcolsep}{3.3pt}
\centering
\begin{tabular}{lllllllll}
\hline\hline

 \noalign{\smallskip}
{Galaxy}           &
{R.A. $^a$}             &
{Dec $^a$}              &
{$\epsilon^b$} & 
{[Fe/H]$^c$}        &
{$r_{\rm h}$ $^b$}   &
{$r_{\rm h}$ $^{Exponential}$ }    &
{$r_{\rm h}$ $^{Plummer}$ }        &
{$r_{\rm h}$ $^{S\acute{e}rsic}$ } \\

 &
(J2000) &
(J2000) &
 & 
(dex) &
(arcmin) &
(arcmin) &
(arcmin) &
(arcmin) \\

\hline
\multicolumn{9}{l}{}\\ 
And\,I $^d$ & 00 45 39.80 & $+38$ 02 28.00 & $0.28\pm0.03$ & $-1.45\pm0.04$ & $3.90\pm0.10$ & $3.20\pm0.30$ & - & - \\
\\
And\,II  & 01 16 29.78 & $+33$ 25 08.75 & $0.16\pm0.02$ & $-1.64\pm0.04$ & $5.30\pm0.10$ & 5.32$^{+0.19}_{-0.02}$ & 5.21$^{+0.09}_{-0.12}$ & 5.35$^{+0.16}_{-0.05}$ \\
\\
And\,III & 00 35 33.78 & $+36$ 29 51.91 & $0.59\pm0.04$ & $-1.78\pm0.04$ & $2.20\pm0.20$ &  2.33$^{+0.07}_{-0.20}$ & 2.21$^{+0.19}_{-0.08}$ & 2.15$^{+0.25}_{-0.02}$  \\
\\
And\,V & 01 10 17.10 & $+47$ 37 41.00 & 0.26$^{+0.09}_{-0.07}$ & $-1.6\pm0.3$  & 1.60$^{+0.20}_{-0.10}$ &  1.88$^{+0.04}_{-0.17}$ & 1.84$^{+0.08}_{-0.14}$ & 1.86$^{+0.06}_{-0.16}$ \\
\\
And\,VI & 23 51 46.30 & $+24$ 34 57.00 & $0.41\pm0.03$ & $-1.3 \pm0.14$ & $2.30\pm0.20$ & 2.24$^{+0.15}_{-0.03}$ & 2.00$^{+0.02}_{-0.16}$ & 2.27$^{+0.12}_{-0.07}$ \\
\\
And\,VII $^e$ & 23 26 31.74 & $+50$ 40 32.57 & $0.13\pm0.04$ & $-1.40\pm0.30$ & $3.50\pm0.10$ & $3.80\pm0.30$ & - & - \\
\\
And\,IX $^f$ & 00 52 53.00 & $+43$ 11 45.00 & 0.00$^{+0.16}_{-0.00}$ & $-2.2\pm0.2$ & 2.00$^{+0.30}_{-0.20}$ & $2.50\pm0.26$ & - & - \\
\\
And\,X & 01 06 33.70 & $+44$ 48 15.80 & 0.10$^{+0.34}_{-0.10}$ & $-1.93\pm0.11$ & 1.10$^{+0.40}_{-0.20}$ & 1.30$^{+0.35}_{-0.20}$ & 1.29$^{+0.36}_{-0.19}$  & 1.28$^{+0.37}_{-0.18}$ \\
\\
And\,XI & 00 46 20.00 & $+33$ 48 05.00 & 0.19$^{+0.28}_{-0.19}$  & $-2.0\pm0.2$ & $0.60\pm0.20$ &  0.54$^{+0.18}_{-0.06}$ & 0.53$^{+0.19}_{-0.05}$ &  0.53$^{+0.19}_{-0.05}$  \\
\\
And\,XII & 00 47 27.00 & $+34$ 22 29.00 & 0.61$^{+0.16}_{-0.48}$  & $-2.1 \pm0.2 $ & 1.80$^{+1.20}_{-0.70}$ & 1.72$^{+0.33}_{-0.18}$ & 1.72$^{+0.33}_{-0.18}$ &  1.73$^{+0.33}_{-0.18}$ \\
\\
And\,XIII & 00 51 51.00 & $+33$ 00 16.00 & 0.61$^{+0.14}_{-20}$ & $-1.9\pm0.2$ & 0.80$^{+0.40}_{-0.30}$ & 0.71$^{+0.25}_{-0.07}$ & 0.69$^{+0.27}_{-0.05}$ & 0.73$^{+0.23}_{-0.09}$ \\
\\
And\,XIV & 00 51 35.00 & $+29$ 41 49.00 & 0.17$^{+0.16}_{-0.17}$ & $-2.26\pm0.05$ & $1.50\pm0.20$ & 1.67$^{+0.20}_{-0.17}$ & 1.66$^{+0.21}_{-0.16}$   & 1.66$^{+0.21}_{-0.16}$ \\
\\
And\,XV & 01 14 18.70 & $+38$ 07 03.00 & $0.24\pm0.10$ & $-1.8\pm0.2$ & $1.30\pm0.10$ & 1.55$^{+0.01}_{-0.25}$ & 1.48$^{+0.08}_{-0.18}$ & 1.46$^{+0.10}_{-0.16}$ \\
\\
And\,XVI & 00 59 29.80 & $+32$ 22 36.00 & $0.29\pm0.08$ & $-2.1\pm0.2$ & $1.00\pm0.10$ & 1.19$^{+0.01}_{-0.19}$ & 1.14$^{+0.06}_{-0.14}$ & 1.13$^{+0.07}_{-0.13}$ \\
\\
And\,XVII  & 00 37 07.00 & $+44$ 19 20.00 & $0.50\pm0.10$ & $-1.9\pm0.2$ & $1.48\pm0.30$ & 1.48$^{+0.01}_{-0.17}$ & 1.47$^{+0.02}_{-0.17}$ & 1.48$^{+0.02}_{-0.17}$  \\
\\
And\,XVIII & 00 02 14.50 & $+45$ 05 20.00 & 0.03$^{+0.28}_{-0.03}$ & $-1.8\pm0.1$ & $0.80\pm0.10$ & 0.97$^{+0.15}_{-0.01}$ & 0.95$^{+0.01}_{-0.15}$ & 0.93$^{+0.03}_{-0.13}$ \\
\\
And\,XIX & 00 19 32.10 & $+35$ 02 37.10 & 0.58$^{+0.05}_{-0.10}$ & $-1.9\pm0.1$ & 14.20$^{+3.40}_{-1.90}$ & 14.28$^{+0.49}_{-0.08}$ & 14.22$^{+0.55}_{-0.02}$ & 14.44$^{+0.33}_{-0.24}$\\
\\
And\,XX & 00 07 30.70 & $+35$ 07 56.40 & 0.11$^{+0.41}_{-0.11}$ & $-1.5\pm0.1$ & 0.40$^{+0.20}_{-0.10}$ & 0.50$^{+0.04}_{-0.10}$ & 0.49$^{+0.05}_{-0.09}$ & 0.46$^{+0.07}_{-0.06}$ \\
\\
And\,XXI & 23 54 47.70 & $+42$ 28 15.00 & 0.36$^{+0.10}_{-0.13}$ & $-1.8\pm0.2$ & 4.10$^{+0.80}_{-0.40}$ & 3.83$^{+0.27}_{-0.55}$ & 3.82$^{+0.28}_{-0.54}$ & 3.82$^{+0.28}_{-0.54}$\\
\\
And\,XXII & 01 27 40.00 & $+28$ 05 25.00 & 0.61$^{+0.10}_{-0.14}$ & $-1.8$ & 0.90$^{+0.30}_{-0.20}$ & 0.90$\pm0.18$ & 0.87$^{+0.21}_{-0.15}$ &  0.85$^{+0.23}_{-0.13}$\\
\\
\hline

\end{tabular}
\vspace{2mm}
\begin{minipage}{0.95\textwidth}
\footnotesize
\textit{Notes.} \\
$^a$ Coordinates inferred from the NASA/IPAC Extragalactic Database (NED)\footnote{\url{https://ned.ipac.caltech.edu/}} portal.\\
$^b$ All ellipticities ($\epsilon$) and half-light radius in the 6$^{th}$ column ($r_{\rm h}$) are referred from \cite{2016ApJ...833..167M}, except for And\,VI which is from \cite{2012AJ....144....4M}. \\
$\epsilon = 1 - b/a$, where $b$ is the semi-minor axis and $a$ is the semi-major axis. \\
$^c$ \cite{2012AJ....144....4M}, $^d$ \cite{2020ApJ...894..135S}, $^e$ \cite{2021ApJ...910..127N}, and $^f$ \cite{2023ApJ...948...63A}. \\
$r_{\rm h}^{\text{Exponential}}$, $r_{\rm h}^{\text{Plummer}}$, and $r_{\rm h}^{\text{S\'ersic}}$ are calculated in this work.\\
And~I, And~VII, and And~IX are not re-analyzed in this study; their half-light radii are adopted from previous literature (\cite{2020ApJ...894..135S, 2021ApJ...910..127N, 2023ApJ...948...63A}) for the method used in those works, while the remaining methods are marked with “–”.
\end{minipage}

\label{table:objects}
\end{table*}

\section{Conclusions}
\label{sec:Conclusions}

In this study, we have identified and characterized LPV stars in 17 satellite galaxies of the Andromeda system, determining their mean $i$- and $V$-band magnitudes as part of a systematic analysis of their photometric and structural properties. From our multi-epoch photometry, we derived the TRGB and half-light radii using different fitting profiles, finding consistent and reliable measurements for all galaxies.

Our TRGB-based distance moduli are in excellent agreement with independent measurements from RR~Lyrae and Cepheid variables, confirming the robustness of our photometric calibration and extinction corrections. The derived half-light radii, presented in Table~\ref{table:objects}, further demonstrate the structural diversity among the M31 satellites, ranging from compact to extended systems and reflecting their varied evolutionary and dynamical histories.

The combination of LPV identification and structural characterization provides a powerful diagnostic of both the stellar content and evolutionary stage of these galaxies. LPVs, as tracers of late stellar evolution, offer unique insights into the ongoing mass-loss processes and dust production that influence the chemical enrichment of the interstellar medium. These results lay the groundwork for reconstructing the star-formation histories and dust feedback of individual systems, thereby contributing to a more comprehensive view of the assembly and evolution of the M31 satellite population.

Future work will expand on these results by incorporating near-infrared photometry and time-series analysis to refine LPV amplitudes, periods, and mass-loss rates, linking the observed variability to detailed stellar evolution models and population synthesis predictions.


\section*{\small Acknowledgements}
\scriptsize{I sincerely thank the organizers at BAO for their kind support and excellent coordination of the meeting, which fostered engaging discussions and fruitful collaborations. I am also grateful to Elham Saremi for her valuable contributions to the observations and data reduction that made this study possible. Support for H.A. was provided through the SeismoLab project (Élvonal grant KKP-137523) of the Hungarian Research, Development and Innovation Office (NKFIH).}

\clearpage 

\scriptsize
\bibliographystyle{ComBAO}
\nocite{*}
\bibliography{ComBAO}

\newpage
\appendix
\renewcommand{\thesection}{\Alph{section}.\arabic{section}}
\setcounter{section}{0}
\normalsize

\end{document}